# The effect of cooling rate on aging in spin glasses

Dinah Parker, François Ladieu, Jacques Hammann, and Eric Vincent

*Service de Physique de l'Etat Condensé (CNRS URA 2464), DSM/DRECAM/SPEC, CEA Saclay, 91191 Gif sur Yvette Cedex, France*

Aging is a well known property of spin glass materials and has been investigated extensively in recent years. This aging effect is commonly observed by thermal remnant magnetization (TRM) experiments in which the relaxation of the magnetization is found to be dependent on the time, $t_w$, spent at constant temperature before a field cut. The TRM curves scale with $t_w^{\mu}$, where $\mu$ is less than 1, which is known as a "subaging" effect. The question of whether this subaging effect is intrinsic, or due to experimental artifacts, remains as yet unanswered. One possible experimental origin of subaging arises from the cooling of the sample to the measuring temperature and it has been proposed that with fast enough cooling $\mu$ would go to 1[1]. Here we investigate this possibility by studying the effect of cooling protocol on aging for 3 well characterized spin glasses, $CdCr_{1.7}In_{0.3}S_4$, Au:$Fe_{8\%}$ and $Fe_{0.5}Mn_{0.5}TiO_3$. We find no strong influence of the cooling rate on $\mu$ and no evidence that $\mu$ would go to 1 for very short cooling times. We propose additionally an argument which shows that small (±150 mK) variations in the temperature of the sample during the first tens of seconds of the TRM can significantly influence the behavior of the relaxation of a spin glass which in turn may result in a misleading interpretation of $\mu$ values.

## 1. Introduction

Spin glasses have been studied for many years and have been found to exhibit a wide range of interesting phenomena arising from slow dynamics[2]. It has been found that the slow relaxation of the magnetization following a change in applied field (known as the thermal remnant magnetization, TRM) is dependent on the waiting time, $t_w$, for which the sample is held at constant temperature before the field change. This is known as an aging effect [3 4 5]. An approximate scaling of a series of TRM curves with different values of $t_w$ can be achieved by plotting $M/M_{FC}$ (the magnetization normalized to the field cooled magnetization) against $t/t_w$. In order to achieve a more precise scaling it is necessary to adjust the waiting time by plotting the magnetization against a scaling variable $\lambda/t_w^{\mu}$ (see details in [2]), where $\lambda \approx t$ for $t \ll t_w$ . TRM studies of spin glass materials have consistently found $\mu < 1$ (typically $0.8 < \mu < 0.9$ [4 5]) and this is referred to as a 'subaging' effect. As $M/M_{FC}$ is dimensionless and $\lambda$ has the dimension of time, the introduction of $\mu < 1$ gives an irrational dimension to the scaling parameter which could be indicative of a hidden timescale. An important question that arises is whether this subaging is an intrinsic property of spin glasses or if it is due to experimental artifacts.

Firstly, we can consider the size limitation of the spin glass samples which are typically polycrystalline solids or powders. The sample can be thought of as being composed of a collection of subsystems, each having an individual finite ergodic time needed to explore all possible configurations. As these times are finite it follows that, over time, they will progressively be reached and $\mu$ will decrease and eventually go to zero when all ergodic times have been satisfied. It is therefore possible that size effects lead to a natural mechanism for sub-aging in spin glasses. In references [6 7] this idea was developed to estimate, using observed values of $\mu$, the cut-off of the barrier height distribution or the size distribution of the considered samples. There is however currently no conclusive experimental proof of this theory.

Another important consideration is the amplitude of the excitation field applied in the TRM procedure. It is currently not clear how, and to what extent, this field perturbs the sample. A systematic study has been performed [8] and it was found that decreasing the size of the excitation field leads to an increase in $\mu$ but that this effect saturates at low fields and $\mu$ remains less than 1. This has been confirmed for $CdCr_{1.7}In_{0.3}S_4$ where $\mu$ remains constant at 0.85 for excitation fields ranging from 10 to $10^{-3}$ Oe [9].

Recently there have been several investigations into how the sample cooling time, which is unavoidably long compared to microscopic time due to the experimental setup, influences aging in spin glasses [1 10 11 12]. If the effect of the (necessarily slow) cooling procedure is to establish a non-negligible age in the initial state obtained after cooling, this yields an underestimate of the $t_w$ values that is numerically equivalent to a decrease of the scaling exponent $\mu$ (this effect may be important for the shortest $t_w$ values). Hence, cooling rate effects may contribute to subaging behavior.

The effect on aging of the thermal history has been the subject of numerous experimental studies [13 14 15], among which “rejuvenation and memory effects” appear as a prominent feature: namely, when a spin glass is cooled step by step, in each new cooling step part of the aging processes restarts (rejuvenation), while the memory of previous agings is retrieved during re-heating. All these results suggest that aging at a lower temperature is hardly influenced by aging at the higher temperatures explored during the cooling procedure. However, when studied in more detail [16], it appears that aging in spin glasses is influenced by both temperature specific processes (rejuvenation and memory effects) and cooling rate effects, as is the case in structural and polymer glasses.

In [10] the relative effects of cooling procedure were evaluated by comparing the relaxation rates of the magnetization following a field change. It was concluded that the relaxation rate was dependent both on $t_w$ and the rate at which the sample was cooled to the measurement temperature, $T_m$. The influence of the cooling rate was most significant for shorter $t_w$ relaxations; for longer values of $t_w$ almost no influence of the cooling rate was observed. It was also found that making a stop lasting 1000 s during the cooling at temperatures close to $T_m$ had a significant influence on the $t_w$ = 0 s relaxation rate, whereas stops at temperatures further from $T_m$ (> 2 K) had no effect. It was therefore concluded that the cooling rate close to $T_m$ influences the relaxation of the magnetization when $t_w$ is short (< 1000 s).

In a recent paper, Rodriguez *et al* [1] investigated aging in a $Cu:Mn_{6\%}$ spin glass in order to determine whether the cooling process is related to the subaging effect. They found that by cooling very quickly from above $T_g$, thereby undershooting the measurement temperature, and then heating back up to the measurement temperature gave a shorter effective cooling than simply cooling directly to the measuring temperature. Four different protocols were used for the study with different cooling rates and undershoot temperatures. An effective cooling time ($t_c^{eff}$) for each protocol was defined as the maximum in the function $S(t) = -dM(t)/[d(log_{10}(t)]$ of the zero waiting time TRM (ZTRM). For different cooling protocols they found $t_c^{eff}$ ranging from 19 to 406 s. An increase in $\mu$ value with decreasing $t_c^{eff}$ was observed, the shortest cooling time giving a $\mu$ of 0.999. It was therefore concluded that $\mu$ would indeed go to 1 for very short cooling times and that subaging is not an intrinsic property of spin glasses but merely an experimental feature.

In this paper we report investigations into the effect of cooling rate on aging in three well characterized spin glasses, $CdCr_{1.7}In_{0.3}S_4$, $Au:Fe_{8\%}$ and $Fe_{0.5}Mn_{0.5}TiO_3$ (compared for instance in [17]). We have based our experimental procedure on the method of Rodriguez *et al* [1]. $Fe_{0.5}Mn_{0.5}TiO_3$ is a monocrystalline sample with very strong uniaxial anisotropy, which is considered to be a good example of an Ising spin glass [18]. $Au:Fe_{8\%}$ and $CdCr_{1.7}In_{0.3}S_4$ are closer to Heisenberg spin glass realizations, with some random anisotropy arising from Dzyaloshinsky-Moriya interactions [19]. Their relative anisotropy constants $(K/T_g)/(K/Tg)_{AgMn}$, measured by torque experiments and normalized to that of $Ag:Mn_{2.7\%}$ [19], are respectively 8.25 ($Au:Fe_{8\%}$) and 5.0 ($CdCr_{1.7}In_{0.3}S_4$). Our motivation for studying these three samples was that we expect cooling rate effects to be greater in the samples of higher anisotropy, as it has been found previously that aging at temperatures close to $T_m$ has a greater influence on the TRM relaxation for more anisotropic spin glasses [17].

## 2. Experimental

All measurements reported in this paper were taken using a Cryogenics S600 SQUID magnetometer. The TRM protocol is as follows: the sample is cooled from a temperature above $T_g$ down to the measuring temperature, $T_m$, under a small excitation field, H. After a waiting time, $t_w$, the field is cut and the relaxation of the magnetization measured over a time, *t*. It is important that the excitation field is small to ensure that the response of the sample remains in the linear regime, thereby avoiding any influence of the field on $\mu$. For all samples investigated in this paper we have used a $T_m$ of 0.7 $T_g$ and an excitation field of 2 Oe for $CdCr_{1.7}In_{0.3}S_4$ and 10 Oe for both $Fe_{0.5}Mn_{0.5}TiO_3$ and $Au:Fe_{8\%}$. This low field range is in the region of linear response for these three samples. The field cooled (FC) and zero field cooled (ZFC) magnetization versus temperature plots for all three samples are shown in figure 1.

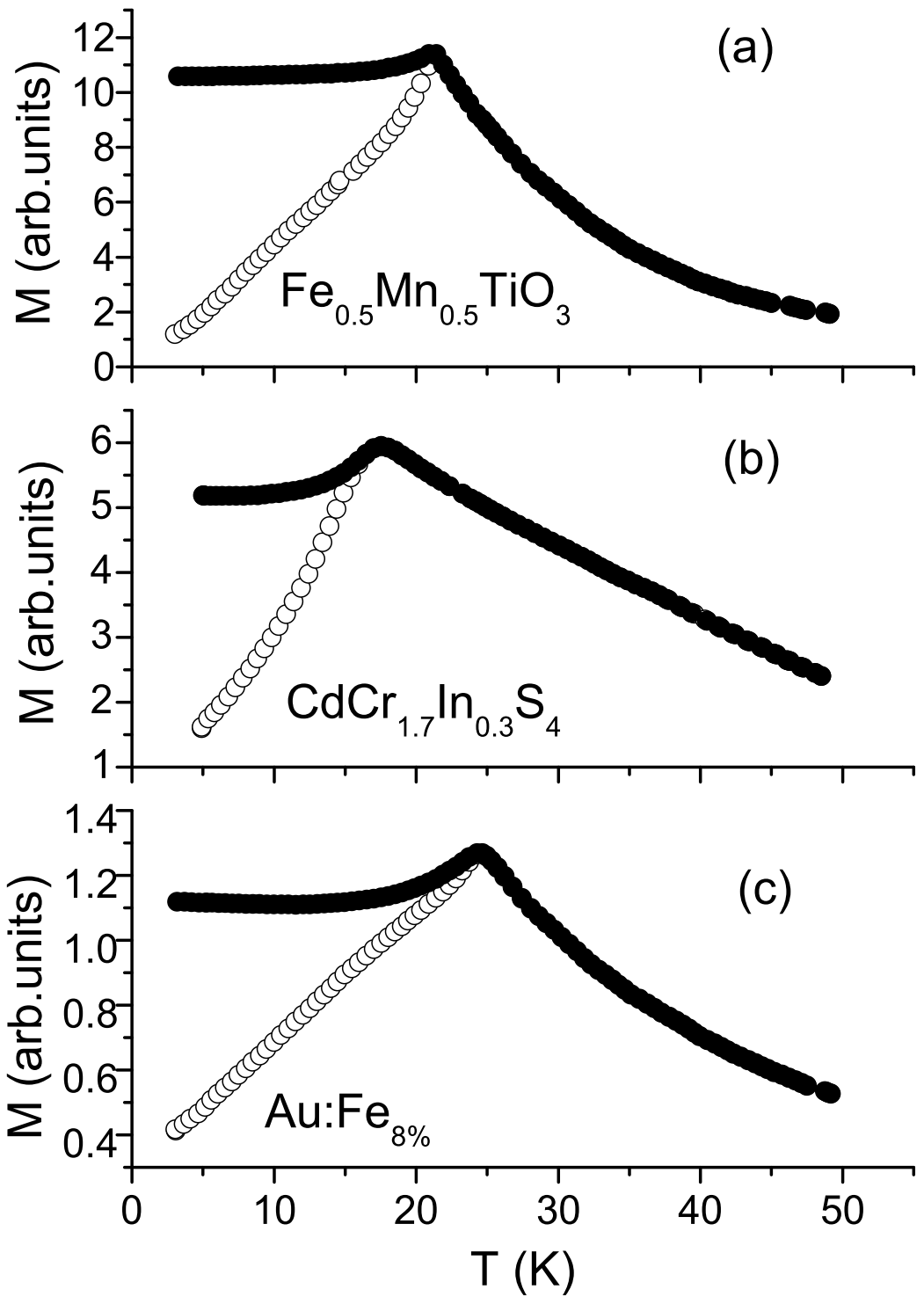

**Figure 1 Field cooled (full circles) and zero field cooled (open circles) magnetization versus temperature curves for (a) $Fe_{0.5}Mn_{0.5}TiO_3$ (H = 10 Oe), (b) $CdCr_{1.7}In_{0.3}S_4$ (H = 2 Oe) and (c) Au:$Fe_{8\%}$ (H = 10 Oe)**

Three different cooling protocols have been employed to investigate the influence of cooling time on $\mu$. These can be evaluated by finding $t_c^{eff}$, the time at which there is a maximum in the relaxation rate, $S(t) = - dM(t)/[d(log_{10}(t)]$, of the $t_w = 0$ s TRM (ZTRM) as presented in the results section.

## 3. Results

### 3.1 Evaluation of the cooling protocols

Figure 2 shows the three different cooling protocols used in this study; these are illustrated by plotting the FC magnetization of $CdCr_{1.7}In_{0.3}S_4$ versus time during the cooling procedure. The FC magnetization varies with temperature in a well defined manner in this temperature range; therefore it is more accurate to use this as a thermometer than the thermometers built into the cryostat as the magnetization reflects the true temperature variation of the sample itself rather than that of the helium flowing in the cryostat. In order to follow the cooling of the sample closely we measured the magnetization by recording the SQUID drift which enables data to be taken every 100 ms.

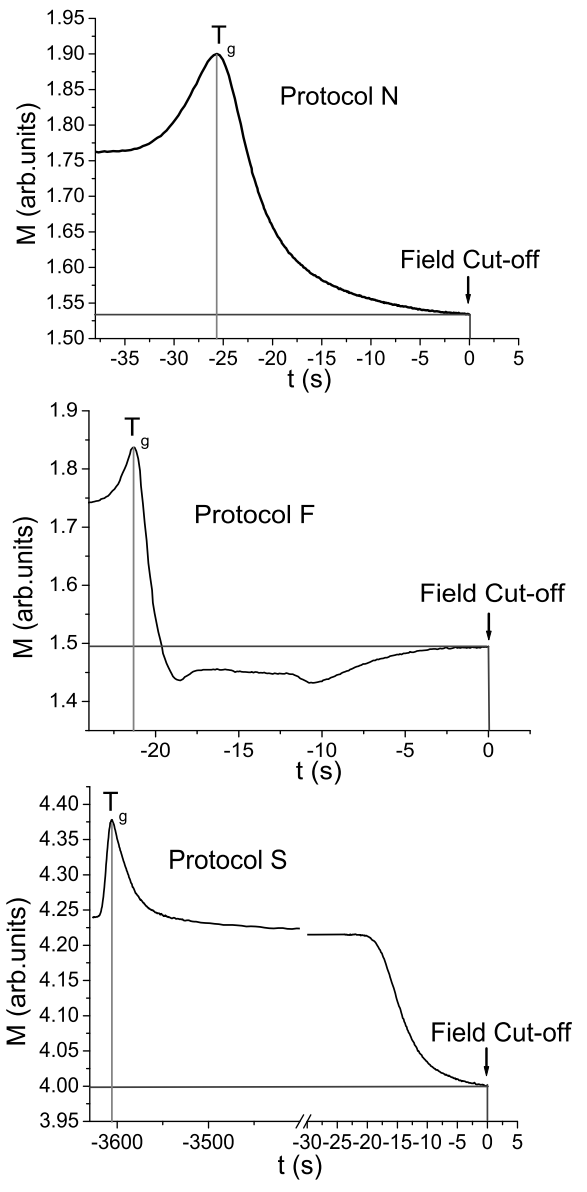

**Figure 2 Cooling protocol N (normal) : direct cooling of the sample from above $T_g$; Cooling protocol F (fast): cooling of the sample to a temperature slightly below $T_m$ and subsequently heating to $T_m$; Cooling protocol S (slow): cooling the sample to a temperature of 0.94 $T_g$, waiting for 1 hr and then cooling directly to $T_m$.**

In protocol N ('normal' cooling) the sample is cooled directly to the measuring temperature, $T_m$, from above $T_g$. This is the standard protocol used in TRM measurements. Protocol F ('fast' cooling) involves cooling the sample at a high rate from above $T_g$ which leads to a temperature undershoot, and then heating the sample back up to $T_m$. These two protocols N and F mirror the methods used by Rodriguez *et al* in [1]. In protocol S ('slow' cooling) the sample is cooled from above $T_g$ to a temperature slightly below $T_g$ ($T/T_g = 0.94$) where the temperature is kept constant for 1 hour after which the sample is cooled to the measuring temperature. We define the start of $t_w$ as the time at which the temperature of the sample, $T$, is $T_m \pm 100$ mK. we have checked that once $T$ enters this temperature interval, it converges towards $T_m$ and does not fluctuate more than +/-10 mK for the whole duration of the TRM experiment. We shall see in section 4 that our definition of $t_w$ leads to a systematic bias $\delta\ t_w$ (in the range of ± 10-20 s) which depends on the type of cooling. The fact that our thermal protocols are not ideal, instantaneous quenches is taken into account by the $t_{ini}$ parameter (see section 3.2). Moreover, except in figure 4 (b), we have disregarded the too short $t_w$ data ($t_w < 200$ s) in accordance with equation (4) of section 4.2.

Values of $t_c^{eff}$ have been extracted for all three cooling protocols for all spin glasses investigated. These values of $t_c^{eff}$ should be considered as a *qualitative* evaluation of the cooling procedures rather than a precise measure of the age of the system after cooling.

In order to carry out an accurate ZTRM measurement we use an excitation field generated by a resistive coil which can be cut almost instantaneously (<100 ms). The maximum field which can be applied using this method is 1.7 Oe.

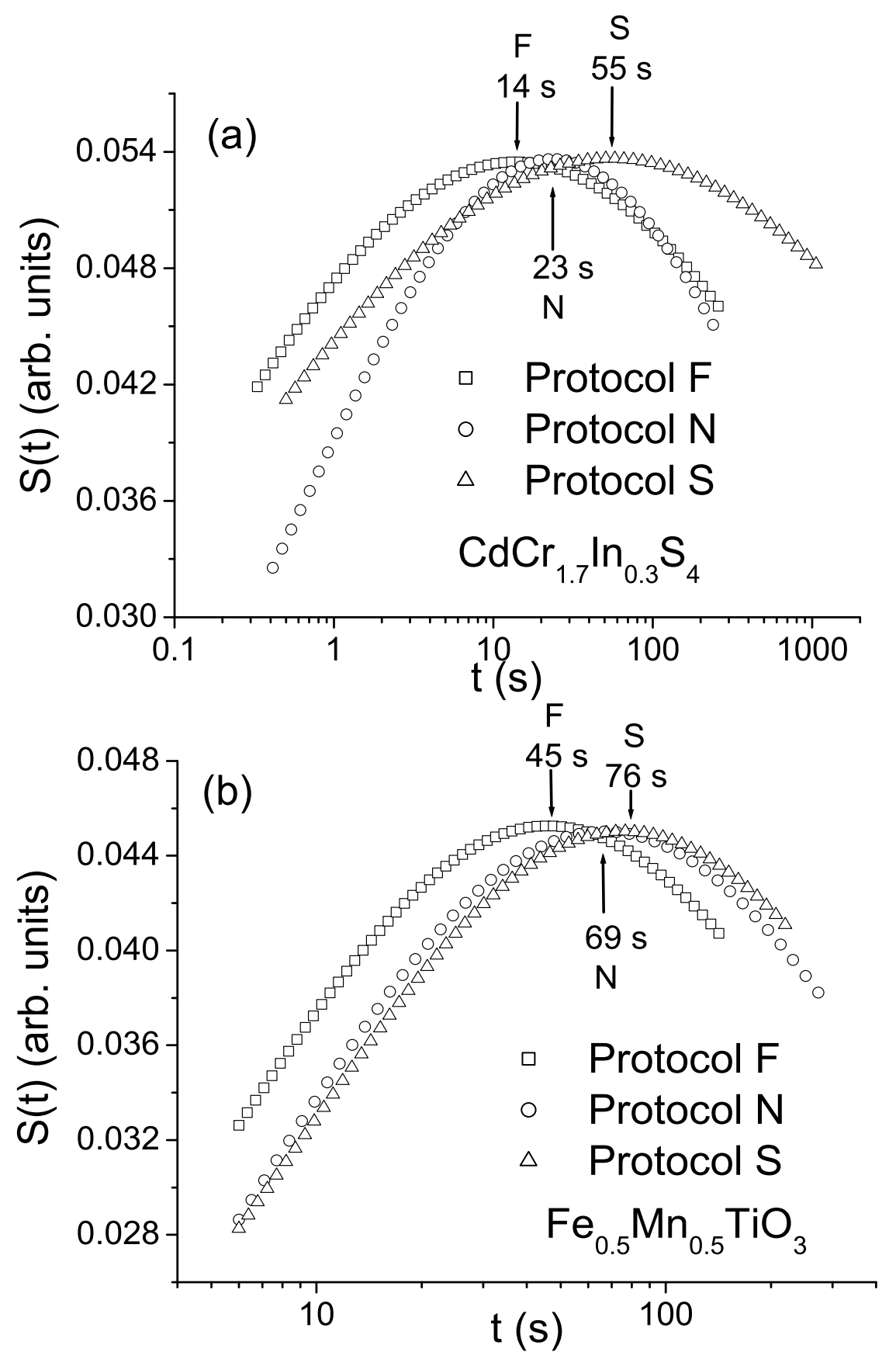

**Figure 3 (a) The relaxation rate S(*t*) versus time of $CdCr_{1.7}In_{0.3}S_4$ measured at 12 K (0.7 $T_g$) after 3 different cooling protocols, N, F and S; (b) The relaxation rate S(*t*) versus time of $Fe_{0.5}Mn_{0.5}TiO_3$ measured at 15.2 K (0.7 $T_g$) after 3 different cooling protocols, N, F and S**

Figure 3 (a) shows S(*t*) versus *t* of the ZTRM curves of all three cooling protocols for $CdCr_{1.7}In_{0.3}S_4$. Here the magnetization was measured by following the SQUID drift which allows us to obtain data for very small values of *t*. Cooling protocols N, F and S give $t_c^{eff}$ values of 23, 14 and 55 s respectively. We therefore conclude that protocol F, with a temperature undershoot, effectively diminishes the effect of the cooling procedure with respect to direct cooling (protocol N). This is consistent with the results of Rodriguez *et al* [1]. As protocol S gives the longest value of $t_c^{eff}$ it seems that waiting for one hour at a temperature slightly below $T_g$ can give an increase in the effect of the cooling procedure with respect to the direct cooling.

The S(*t*) versus *t* curves of $Fe_{0.5}Mn_{0.5}TiO_3$ resulting from all three cooling protocols are shown in figure 3(b). In this case it was not possible for us to measure the magnetization using the SQUID drift method as the data obtained was very noisy due to the weak signal we obtain for $Fe_{0.5}Mn_{0.5}TiO_3$ when using a small (1.7 Oe) excitation field. Instead we have used an oscillating measurement function which enables us to take a data point approximately every 5 seconds. As for $CdCr_{1.7}In_{0.3}S_4$, we find $t_c^{eff}$(protocol F) < $t_c^{eff}$(protocol N) < $t_c^{eff}$(protocol S). The values of $t_c^{eff}$ are larger than those found for $CdCr_{1.7}In_{0.3}S_4$; this is to be expected as it has been found that the higher the spin anisotropy, the greater the influence on aging at $T_m$ by time spent at $T_m \pm \Delta\ T$.[13]

For both $CdCr_{1.7}In_{0.3}S_4$ and $Fe_{0.5}Mn_{0.5}TiO_3$ we find values of $t_c^{eff}$ in the same range as those reported by Rodriguez *et al* [1] in their investigation. It was not possible to carry out the same $t_c^{eff}$ evaluation for Au:$Fe_{8\%}$ as the signal from this sample was too weak to give smooth data even when using the oscillating method to measure the magnetization. The normal extraction procedure, which was used for the $t_w \neq 0$ TRM measurements reported below in this paper, only allows one data point to be measured every ≈ 25 s and is therefore not suitable for the investigation of the ZTRM curves where we expect to find an inflection point at $t < 25$ s.

## 3.2 Influence of the cooling time on aging

The TRM curves for $CdCr_{1.7}In_{0.3}S_4$ following the cooling protocols N, F and S are shown in Figure 4. Figures 4 (a), (b) and (c) show the curves plotted against $t/t_w$ and it is clear in all cases that the curves do not scale well. Figures 4 (d), (e) and (f) show the scaling of the TRM curves using the standard scaling procedure for spin glasses.[2, 20] The $t_w$ independent term $A(\tau_0/t)^{\alpha}$ is subtracted from the total normalized magnetization ($M/M_{FC}$) to account for the stationary part of the TRM; the same values of $A$ and $\alpha$ have been used for all three sets of TRM data. As in references [1, 2], the remaining magnetization, ($M/M_{FC}$) - $A(\tau_0/t)^{\alpha}$, is the aging (non-stationary) part (see however [21]). It is plotted against the scaling variable $\lambda/t_w^{\mu}$, defined as $\lambda/t_w^{\mu} = t_w^{1-\mu}\ [(1+t/t_w)^{1-\mu} -1]/[1-\mu]$. $\lambda$ is an effective time which accounts for the evolution of the aging dynamics during the relaxation and has two simple asymptotic limits: for $t \ll t_w$ , $\lambda/t_w^{\mu} \sim t/t_w^{\mu}$, and for $t \gg t_w$ , $\lambda/t_w^{\mu} \propto (t^{1-\mu}- t_w^{1-\mu})$ (see details in [2, 20]).

For cooling protocol N (normal cooling) we find a $\mu$ of 0.87 which is consistent with previously reported $\mu$ values for $CdCr_{1.7}In_{0.3}S_4$.[22] The results from protocol F, slightly less accurate due to a narrower range of $t_w$, yield $\mu = 0.85$. This change in $\mu$ is small enough that we can consider that, within error, there has been no significant change in $\mu$ with cooling rate. For the TRM curves obtained following the cooling protocol S (slow cooling) we find a $\mu$ value of 0.87 (the same as for normal cooling), however, in this case we found that scaling of the shorter $t_w$ curves could only be achieved by adding an extra time, $t_{ini}$, of 130 s to all $t_w$ values. This $t_{ini}$ value of 130 s is of the same order as the value of $t_c^{eff}$ found for this cooling protocol (55 s). We conclude therefore that the only visible effect of a slower cooling procedure is to establish an initial age, $t_{ini} \approx 100$ s, and that there is no clear trend towards $\mu = 1$ with shorter cooling times in $CdCr_{1.7}In_{0.3}S_4$.

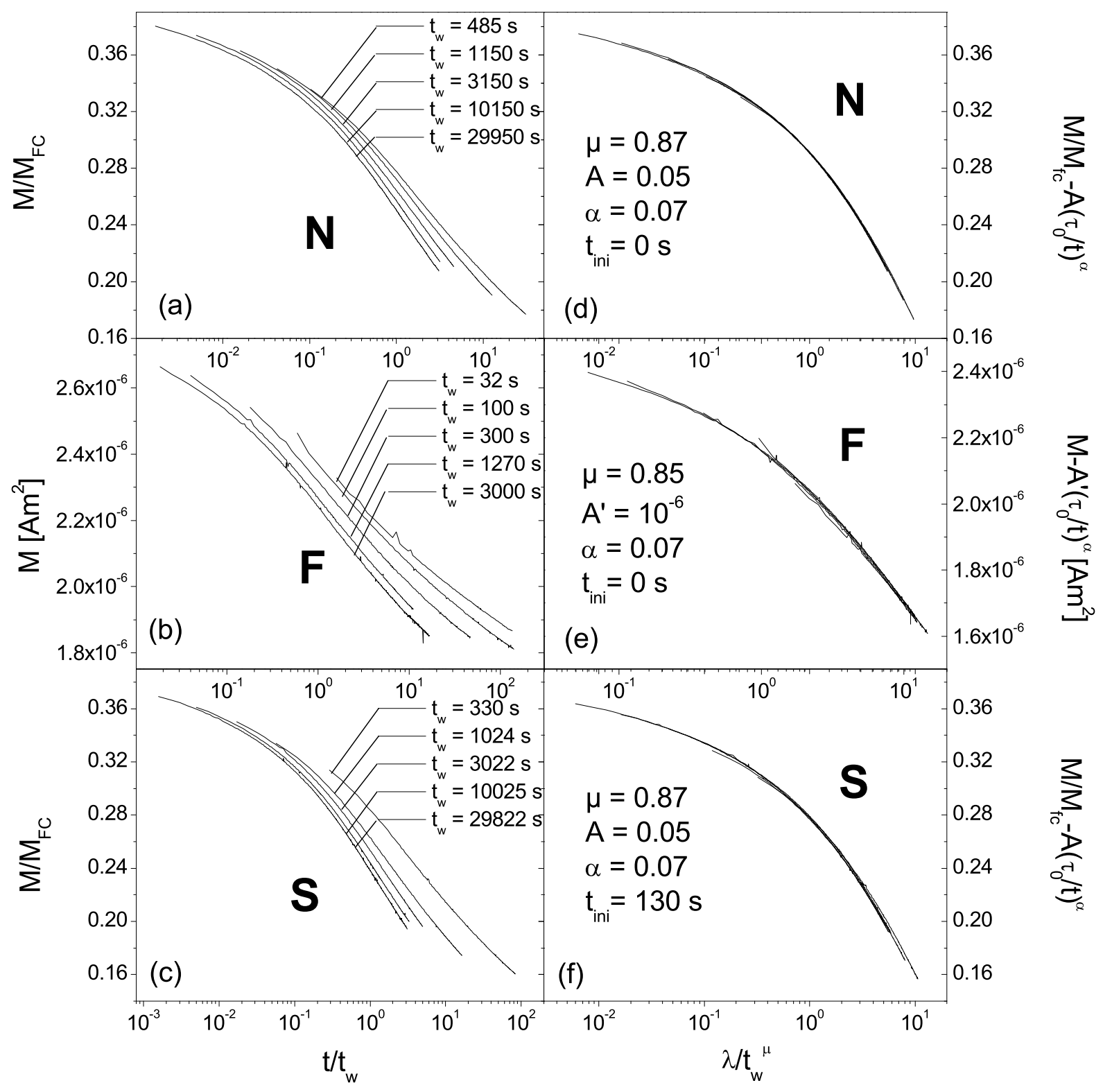

**Figure 4 (left) TRM curves of $CdCr_{1.7}In_{0.3}S_4$ following cooling protocol N (a), protocol F (b) and protocol S (c) plotted against $t/t_w$; (right) scaling of the TRM curves following cooling protocol N (d), protocol F (e) and protocol S (f). Note that in the scaling procedure $t_w$ is in fact $t_w + t_{ini}$ (see text for details). For the protocol F data the exact vertical scale is unknown, but within error $A'$ corresponds to $A$ = 0.05.**

Figure 5 shows the TRM curves for $Au:Fe_{8\%}$ using the three different cooling protocols: (a), (b) and (c) show the curves plotted against $t/t_w$ and (d), (e) and (f) show the scaled curves. As for $CdCr_{1.7}In_{0.3}S_4$, values of $A$ and α are kept constant for all three sets of TRM curves. Cooling protocol N (normal cooling) gives a $\mu$ of 0.86 which is consistent with previous values obtained for $Au:Fe_{8\%}$. With a shorter effective cooling time (protocol F, fast cooling) we see an increase in the $\mu$ value to 0.91. However, cooling protocol S (slow cooling) gives a $\mu$ value of 0.89, greater than that found for normal cooling and therefore there appears to be no clear trend in the variation of $\mu$ with cooling protocol.

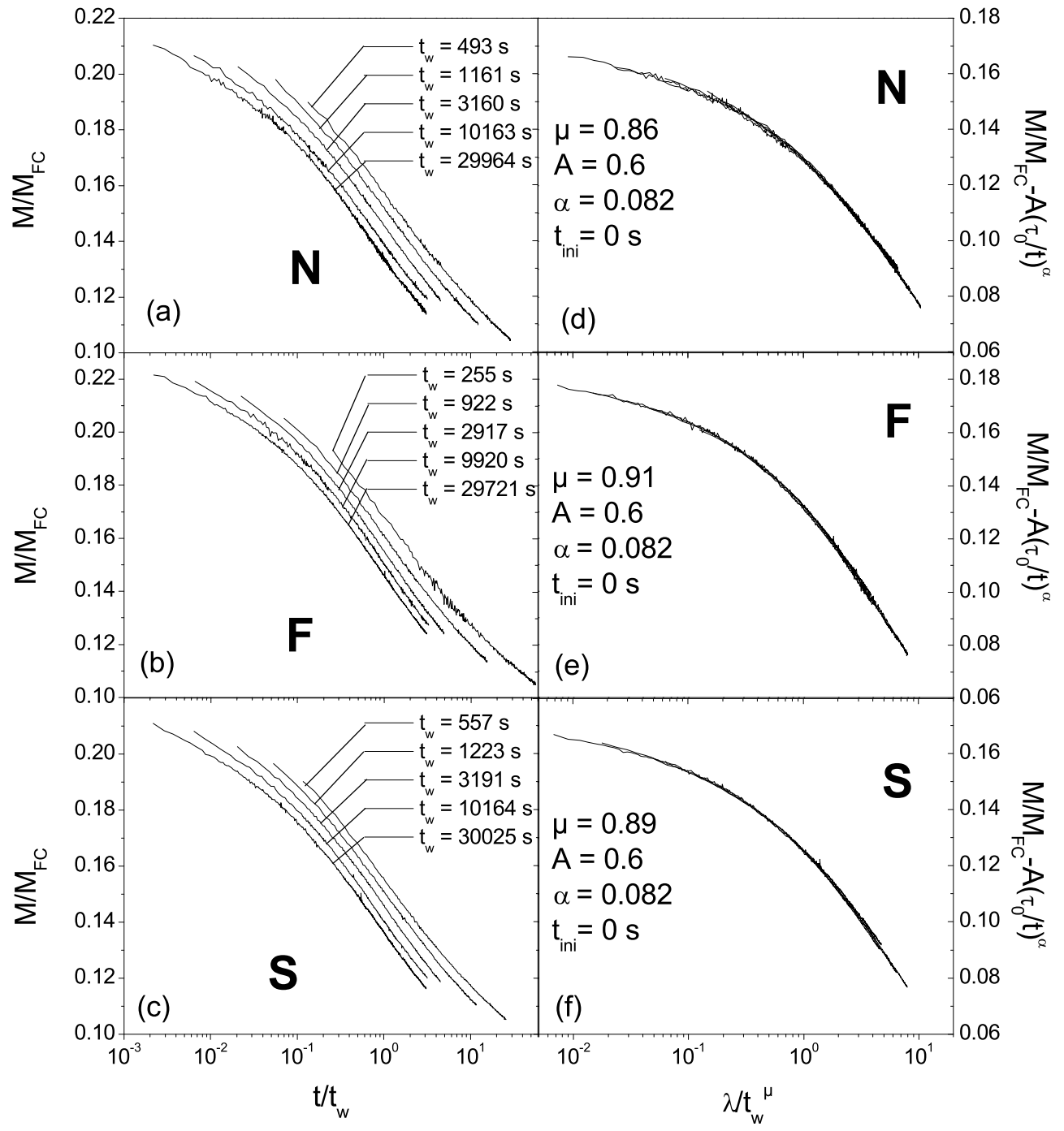

**Figure 5 (left) TRM curves of Au:Fe$_{8\%}$ following cooling protocol N (a), protocol F (b) and protocol S (c) plotted against *t*/*t*$_w$; (right) scaling of the TRM curves following cooling protocol N (d), protocol F (e) and protocol S (f).**

Figure 6 (a) and (d) show the TRM curves of $Fe_{0.5}Mn_{0.5}TiO_3$ for the F and S cooling protocols respectively, plotted against $t/t_w$. Figures 6 (b), (c), (e) and (f) show the scaling of the TRM curves for all three cooling protocols; $A$ and $\alpha$ values have been kept constant in all cases.

For the scaling of the TRM following cooling protocol F (fast cooling, shown in figure 6 c) we find a $\mu$ of 0.84. We find that in order to achieve good scaling of the shorter $t_w$ curves it is necessary to add an extra time, $t_{ini}$ of 75 seconds to all values of $t_w$. For both cooling protocol N (normal cooling) and cooling protocol S (slow cooling) scaling can be achieved in two ways; either by using a $\mu$ of 0.77 with $t_{ini}$ of 75 s (figures 6 b and e), or by using a $\mu$ of 0.84 (as found for cooling protocol F) and adding a $t_{ini}$ of 300 s to all $t_w$ values (figure 6 f, results for protocol N not shown). For both values of $\mu$, we observe a downgrading in the scaling quality for the N and S cooling protocols w.r.t. cooling protocol F. Thus the comparison of the F and N, S protocols in $Fe_{0.5}Mn_{0.5}TiO_3$ shows some influence of a slowing down in the cooling procedure and this influence can be interpreted either by a decrease in $\mu$ or by in increase in $t_{ini}$. The downgraded quality of the scaling does not allow us to distinguish between these two possibilities. In any case, there is no trend of $\mu \rightarrow 1$ for short cooling times.

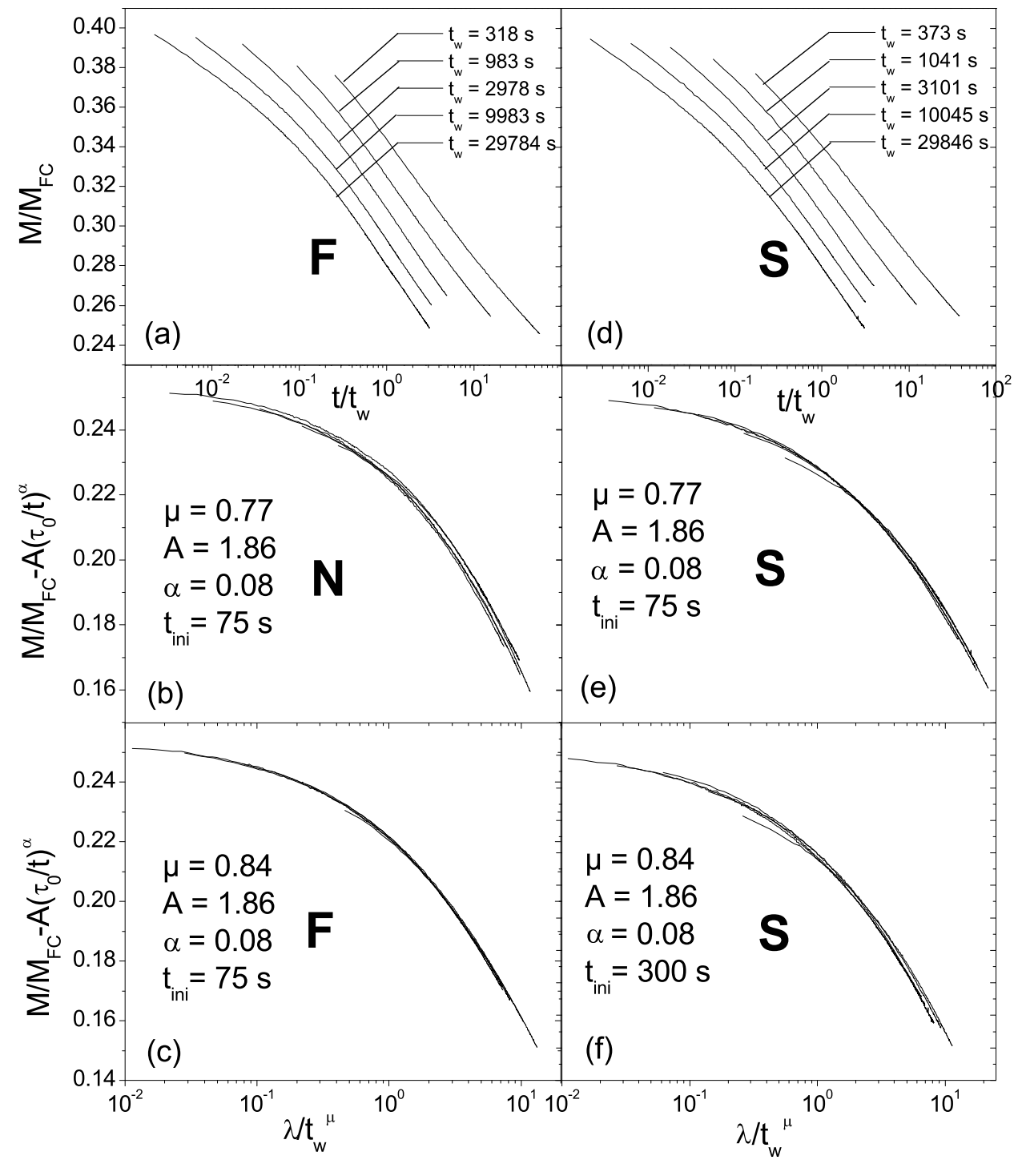

**Figure 6 (a) & (d) TRM curves of $Fe_{0.5}Mn_{0.5}TiO_3$ plotted against $t/t_w$ following cooling protocols F and S respectively; (b) & (c) scaling of the TRM curves following cooling protocols N and F respectively; (e) & (f) scaling of the TRM curves following cooling protocol S (with alternative scaling parameters; see text for details). Note that in the scaling procedure $t_w$ is in fact $t_w + t_{ini}$ (see text for details)**

Therefore, among the three representative spin glass samples that we have investigated here, we find that the effect of the cooling procedure on the scaling parameters remains very weak. A slower cooling may yield a higher value of the initial age, $t_{ini}$ ($Fe_{0.5}Mn_{0.5}TiO_3$, $CdCr_{1.7}In_{0.3}S_4$) or a slightly lower value of $\mu$ ($Fe_{0.5}Mn_{0.5}TiO_3$) but in any case we do not find any sign of a possibility that $\mu$ goes to 1 for shorter cooling times in the limit of experimental constraints imposed by standard cryostats.

## 4. Estimate of the influence of a temperature under/overshoot in the various cooling procedures

## 4.1 Origin of the effect.

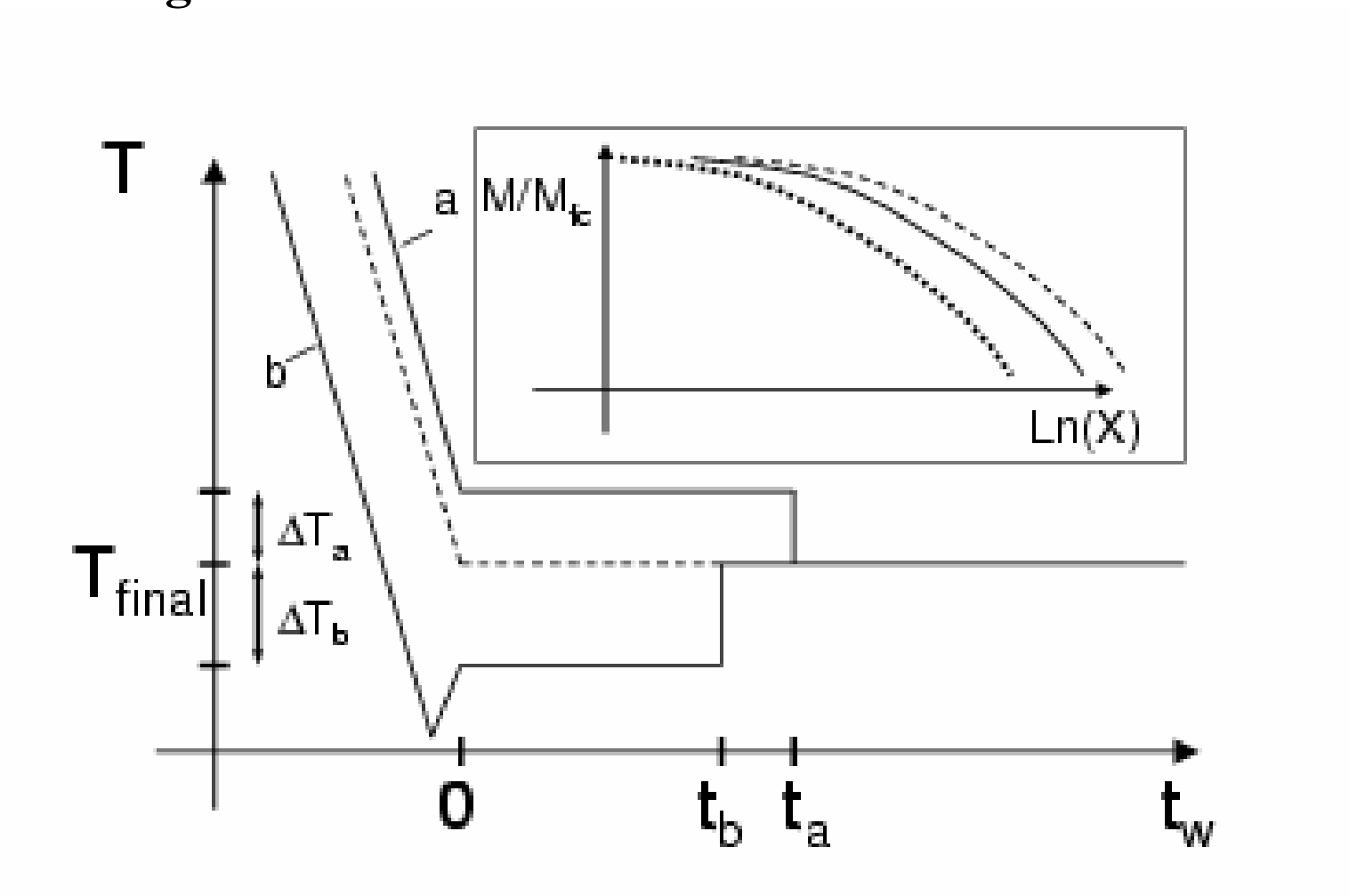

**Figure 7 Schematic diagram of the thermal profile during the first tens of seconds of a TRM experiment (see text for details). The unavoidable slow approach to $T_m$ produces a systematic error on $t_w$ values. In the case "*b*" where $T_m$ is attained "from below", $t_w$ values are overestimated by an amount $\delta t_{wb}$ proportional to $t_b$. Inset: in the scaling plane (*X,M/Mfc*), this will translate the TRM curves by an amount ~ $\delta t_{wb}/t_w$ with respect to the ideal situation where $T_m$ is attained directly (dashed line). The data corresponding to the shortest waiting time $t_{w1}$ will be translated by a greater amount (dotted line) than those corresponding to the longest waiting time $t_{w2}$ (solid line). This will lead to an artificial increase of μ to restore the scaling.**

From our initial analysis we find no clear effect of cooling rate on the value of $\mu$ for any of the three samples studied, which is in contradiction with the results of other authors.[1] In order to address this discrepancy and to analyze our results in more detail, we shall now present an argument that takes into account the fine details of the cooling protocols used.

The basis of our idea is that, as spin glasses are extremely sensitive to the temperature, *T*, one has to pay great attention to the details of the thermal history during the first tens of seconds of the waiting time, $t_w$, of a TRM. Any deviations from the measuring temperature during this time will be especially significant for short $t_w$ TRM curves.

As illustrated in figure 7, two main possibilities can be considered concerning the precise value of *T(t)* in the first tens of seconds of $t_w$. Due to the limits of the experimental setup the temperature will never stabilize at $T_m$ at exactly $t_w = 0$ s and therefore, as illustrated schematically in figure 7, during the first tens of seconds of $t_w$ the sample will either be at a temperature slightly above ($\Delta T_a$) or slightly below ($\Delta T_b$) $T_m$. We will consider precise values of $\Delta T_a$ and $\Delta T_b$ in section 4.2. In this $t_{a,b}$ interval *T(t)* is very close to $T_m$ and will therefore contribute to aging, however this contribution will be different for cases *a* and *b*.

The effect of these unavoidable $\{t_{a,b};\ \Delta T_{a,b}\}$ can be estimated from previous works *e.g.* [17] [23] where the effects of temperature variations, $\Delta T$, during $t_w$ have been studied in detail. Spending a time $t_2$ at a temperature $T_m - \Delta T$ gives rise to a TRM relaxation that amounts to isothermal aging during a smaller effective time $t_{2eff} < t_2$. This means that when $T_m$ is approached "from below", the timed $t_w$ values will overestimate the "real" $t_w$ values (which we will call $t_w$*) by a systematic positive amount $\delta t_{wb}$. For a Heisenberg-like spin glass with $\Delta T_b$= -0.15 K = -0.005 $T_g$ and $t_b$ = 50 s, we get, by using references [17 23], $\delta t_{wb} \approx$ +15 s.

For positive $\Delta T$s one finds $t_{2eff}$ larger than $t_2$ (see [24]) provided $\Delta T/T_g$ is small enough. The timed $t_w$ values will thus underestimate the $t_w$* values by an amount $\delta t_{wa}$. Getting a precise value of $\delta t_{wa}$ is less straightforward than for $\delta t_{wb}$, due to the fact that the "cumulative aging" only holds for very small positive $\Delta T$s (for larger $+\Delta T$s the shape of the final TRM is no longer that of an isothermal TRM). With $\Delta T_a$ = +0.15 K and $t_a$ = 30 s, we shall take $\delta t_{wa} \approx$ -15 s as a low estimate.

### 4.2 Quantitative evaluation of the influence of the cooling procedure

As the orders of magnitude of $\delta t_{wa,b}$ are not negligibly small with respect to the shortest $t_w$ used in experiments, we now move to a quantitative estimate of the effect of $\delta t_{wa,b}$ on the determination of $\mu$. We first define an 'ideal' case of a direct cooling to $T_m$ with instant temperature stabilization where $t_w$* are the waiting times and $\mu$* the corresponding $\mu$ value which yields a scaling of the aging part of $M/M_{FC}$.

In a real experiment where $T_m$ is approached from below, we have $t_w = t_w^* + \delta t_{wb}$. Let us first consider the case where the equilibrium part, $A(t_0/t)^{\alpha}$, of $M/M_{FC}$ is negligible. In this case, denoting $X = \lambda/t_w^{\mu}$, the only effect of adding $\delta t_{wb}$ will be a horizontal translation of the data in the ($X$, $M/M_{FC}$) plane by an amount:

$$\delta_{\delta\, tw} \ln(X) \equiv \frac{\partial \ln(X)}{\partial t_w}\delta\, t_{wb} = \frac{\delta\, t_{wb}}{t_w}\left[1-\mu^* - \frac{t}{t_w}\left(\frac{1}{(1+t/t_w)^{\mu^*}\ln(1+t/t_w)}\right)\right] \qquad (1)$$

The corresponding translations are shown schematically in the inset of figure 7. With respect to the ideal (dashed) scaling curve, the data are translated by an amount that increases as $t_w$ decreases, *i.e.* the shift is much greater for the short $t_w$ curves w.r.t. the long $t_w$ curves. The scaling is thus destroyed, as depicted in the inset of figure 7, and we shall see that varying $\mu$ away from its (ideal) $\mu$* value restores the scaling.

To compute this artificial $\mu$ evolution we simplify equation (1) , noting that, since $\delta t_{wb} >$ 0, these $t_w$ dependent translations always shift the curves to lower values.

$$\delta_{\delta tw} \ln(X) \equiv -N\frac{\delta t_{wb}}{t_w} \qquad (2)$$

N is always very close to 1 in practice, evolving from $\mu*$ to ~ $\mu*$ - 0.5 when $t$ goes from << $t_w$ to $t/t_w$ = 10. We observe (see figure 7) that the translations have driven the short $t_w$ curves below the long $t_w$ curves. Hence scaling will be restored by increasing $\mu$ by an amount $\delta\mu = \mu - \mu*$ chosen to compensate the translations described by equations (1) and (2). We define, as in equation (1), the translation due to a $\mu$ shift by:

$$\delta_\mu \ln(X) \equiv \frac{\partial \ln(X)}{\partial \mu}\delta\mu = \delta\mu\left[-\ln(t_w) + \frac{1}{1-\mu} - \frac{\ln(1+t/t_w)\,(1+t/t_w)^{1-\mu}}{(1+t/t_w)^{1-\mu}-1}\right] \cong -\delta\mu \ln(t+t_w) \quad (3)$$

The $\delta\mu$ values are obtained by requiring that the $\delta\mu$ translations compensate the $\delta t_{wb}$ translations. More precisely, we require that the compensation occurs between the differences of the translations between the two curves corresponding to the lowest ($t_{w1}$) and highest ($t_{w2}$) $t_w$ values. We thus obtain $\delta\mu$ by:

$$\left[\delta_\mu \ln(X)\right]_2^1 + \left[\delta_{\delta\, tw} \ln(X)\right]_2^1 = 0 \qquad (3a)$$

which gives:

$$\delta\mu = \left[\frac{N(t_1/t_{w1},\mu*)}{\ln\left(\frac{t_{w,2}(1+t_2/t_{w,2})}{t_{w,1}(1+t_1/t_{w,1})}\right)}\right]\frac{\delta t_{w,b}}{t_{w,1}} \cong \frac{0.75*\delta t_{w,b}}{t_{w,1}\ln\left(\frac{t_{w,2}}{t_{w,1}}\right)} \qquad (4)$$

In equation (4), the first equality was obtained by neglecting the second term $\delta t_{wb}/t_{w2}$ due to the fact that in experiments $t_{w2} >> t_{w1}$. To get the second equality, as $\mu$ is close to 1 in practice, we have replaced $X(t_1,t_{w1}) = X(t_2,t_{w2})$ with $\ln(1 + t_1/t_{w1}) = \ln(1 + t_2/t_{w2})$ without noticeable error. The fact that equation (4) is $t_i$ independent proves that the various $t_{wi}$ dependent translations produced by a given $\delta t_{wb}$ can indeed be compensated almost perfectly, in the overall range of $X$, by a $\delta\mu$ shift.

We note that the value of $\ln(t_{w,2}/t_{w,1})$ in equation (4) is essentially the same in all experiments, as $t_{w,2}$ is always much larger than $t_{w,1}$. This is not the case for $t_{w,1}$, the value of which varies widely between the various studies. Equation (4) demonstrates that choosing a small $t_{w,1}$ (i.e. including small $t_w$ data in the scaling) makes the determined $\mu$ extremely sensitive to the choice of the origin of times chosen for $t_w$. In other words, with a small $t_{w,1}$ , $\mu$ is no longer uniquely defined. The only way to suppress its $\delta t_{wa,b}$ dependence is to discard the small $t_w$ data from the scaling, i.e. to have a $t_{w,1}$ "large enough", according to equation (4).

Let us now move to a quantitative check of equation (4). Looking at the results for $Fe_{0.5}Mn_{0.5}TiO_3$ from figure 6(e)-(f) we observe that increasing $t_{ini}$ from 75 s to 300 s (hence increasing all $t_w$ values by 225 s) gives an increase in $\mu$ of 0.07. This is compatible with the value of $\delta\mu$ = +0.10 that we obtain from equation (4) using $t_{w1}$ = 373 s, $t_{w2}$ = 29848 s and $\delta t_w$ = +225 s.

We shall now use equation (4) to see if it is possible to account for the discrepancies between the experimental results we report in this paper and those of Rodriguez *et al* who concluded that $\mu$ goes to 1 for faster cooling rates. In their investigation, the 'fast' cooling protocols involve a temperature undershoot (*i.e.* with $T_m$ being approached 'from below' during the final stages of cooling) whereas for the 'slow' cooling protocols, cooling was 'from above'. Therefore the higher values of $\mu$ were found for positive $\delta t_{wb}$ and lower values of $\mu$ for negative $\delta t_{wb}$ as predicted qualitatively by equation (4). In order to carry out a quantitative analysis we have conservatively estimated (from figure 1 of [1]) a $\delta t_{wb}$ of +15 s for the 'fast' cooling protocols and –15 s for the 'slow' cooling protocols; $t_{w1}$ = 50 s and $t_{w2}$ = 10000 s. This leads to an artificial increase in $\mu$ of 0.04 for the fast cooling protocols and a decrease in $\mu$ of 0.04 for the slow cooling protocols, giving a total change in $\mu$ of 0.08. This compares well with their experimental observations of a total change in $\mu$ of 0.12.

We emphasize that in Rodriguez's experiments the $\mu$ = 0.999 value is obtained when $T_m$ is attained "from below" where equation (4) predicts an increase in $\mu$. As the authors themselves note, it is the set of curves with low $t_w$ values (in the range [50 s, 1000 s]) which drive $\mu$ to high values: this is directly explained by equation (4). In other works [2 22], the minimum value of $t_w$ is usually in the 300 - 500 s interval. Here, equation (4) predicts that the total $\mu$ excursion should be much smaller for longer $t_w$ curves, below 0.01, *i.e.* hardly observable.

Up to now, we have disregarded the influence of the equilibrium term $A(\tau_0/t)^\alpha$ on the experimental determination of $\mu$. In the appendix we show quantitatively that for some spin glasses, such as Cu:$Mn_{6\%}$ [1] or $CdCr_{1.7}In_{0.3}S_4$, the A parameter is sufficiently small to make the above analysis valid. Moreover we show that A and $\mu$ do not play similar roles in the scaling: a variation of A cannot be compensated by a variation of $\mu$, as has already been shown in [2].

## 5. Conclusion

We have investigated experimentally the TRM of three well characterized spin glass samples, $CdCr_{1.7}In_{0.3}S_4$ (Heisenberg-like), Au:$Fe_{8\%}$ (Heisenberg-like) and $Fe_{0.5}Mn_{0.5}TiO_3$ (Ising), using three different cooling protocols in order to evaluate the influence of the cooling rate on the scaling parameter, $\mu$. The experimental results we report here show no clear trend in $\mu$ with cooling rate, in particular there is no evidence to suggest that $\mu$ goes to 1 with very fast cooling in the limit which can be explored with common experimental procedures. We have proposed a mechanism whereby the values of $\mu$ may be influenced significantly by subtle temperature variations during the first few tens of seconds of $t_w$, which are in turn influenced by the cooling protocol. These changes in $\mu$ are greater when shorter $t_w$ relaxation curves are included in the scaling and are also expected to be greater for Ising as compared to Heisenberg spin glasses. We propose that these effects can account for the differences between the results we report here and those previously reported by Rodriguez *et al* [1] in which it was concluded that $\mu$ goes to 1 with increasing cooling rate.

The origin of $\mu$ less than 1 (and hence the origin of the hidden time scale in the scaling variable $\lambda/t_w^{\mu}$) still remains unresolved. As expressed previously [11], this missing scale might be related to the fact that the state from which aging is started is not random. Indeed, due to the finite amounts of time spent at all the temperatures between $T_g$ and the working temperature, $T$, and despite the rejuvenation effects some spin-spin correlations have already taken place when the sample reaches $T_m$. On the one hand one may argue that these correlations should be short-ranged with respect to those which develop on the long time scales probed during TRM experiments, and thus they can be disregarded. On the other hand, experiments by Zotev *et al* [11] show that the nature of the correlations present in the initial state do have an effect since $\mu$ is greater than 1 when the initial state is zero field cooled (IRM experiments). Moreover, the numerical simulations of Berthier and Bouchaud [25] have shown that some significant decrease in $\mu$ is observed when moving from an infinitely fast quench to a quench whose duration is $\tau_0$. However, investigating this phenomenon further would require either extending the numerics towards macroscopic time scales or finding a way to produce ultrafast experimental quenches. As far as we know, this is at present far from reach.

## Appendix

**The role of the A parameter**

We now briefly expand on the previous arguments to estimate the influence that the equilibrium term $A(\tau_0/t)^{\alpha}$ may have on the experimental determination of $\mu$. Instead of the previous horizontal translations whose magnitude was essentially driven by the $t_w$ value throughout the $X$ available experimental range, we show that we now have to consider vertical translations whose magnitude is strongly dependent on $X$. We start from the case $A = 0$ where $M/M_{FC}$ can be described by:

$$\frac{M}{M_{FC}} = C - p\ln(X) \qquad (5)$$

where $C$ is a constant and $p$ is the magnitude of local slope of $M/M_{FC}$ in the ($\ln(X)$; $M/M_{FC}$) plane (note that $p$ strongly depends on $X$, see below). For a finite $A$, we have to solve the equation:

$$C - p\ln(X(t_1,t_{w1})) - A(\tau_0/t_1)^{\alpha} = C - p\ln(X(t_2,t_{w2})) - A(\tau_0/t_2)^{\alpha} \qquad (6)$$

which has no general analytical solution. However, starting from the case where $A = 0$ where the solution is $X(t_1,t_{w1}) = X(t_2,t_{w2})$, we can treat the case of small $A$s perturbatively, which yields:

$$\left[\delta_A \ln(X)\right]_2^1 = -\frac{A}{p}\left(\frac{\tau_0}{t_1}\right)^{\alpha}\left(1-\frac{t_1^{\alpha}}{t_2^{\alpha}}\right) \qquad (7)$$

Since the right hand side of equation (7) is always negative, we conclude that going from $A = 0$ to $A$ qualitatively increases $\mu$, as previously noted in [2]. Using the same principles as in equation (3a), we obtain:

$$\delta\mu = \frac{A\left(\frac{\tau_0}{t_1}\right)^{\alpha}\left(1-\left(\frac{t_1}{t_2}\right)^{\alpha}\right)}{p\ln\left(\frac{t_{w,2}+t_2}{t_{w,1}+t_1}\right)} \cong \frac{A\left(\frac{\tau_0}{t_1}\right)^{\alpha}\alpha}{p} \qquad (8)$$

where, apart from the fact that $\alpha \ll 1$, we used in the last approximate equality that, since $A$ is small, two points meet in the scaling curve when $X(t_1,t_{w1}) \approx X(t_2,t_{w2})$, which, as in equation (4), amounts to $t_1/t_{w1} \approx t_2/t_{w2}$ when $\mu$ is not far from 1, as in experiments.

From equation (8), we see that a finite $A$ does not really correspond to a simple $\delta\mu$ shift, since $p$ is strongly $X$ dependent. We therefore arrive at the well known conclusion that $A$ and $\mu$ do not play similar roles[2].

In order to confirm this, we can evaluate equation (8) by looking at the experimental scaling curves, where we find that $p$ typically ranges from 0.004 to 0.04 in the $X$ interval [0.01;3] where all the $t_w$ data are included in the scaling. Setting $t_1 = 1$ s (the minimum experimental value), we get (with typical values of $\tau_0 = 10^{-12}$ s and $\alpha = 0.05$) $|\delta\mu| \leq A$. We emphasize that this order of magnitude estimate only holds when $A$ can be treated perturbatively. From equation (7) and using values of $p$ extracted from experimental scaling curves, we deduce that this typically corresponds to $A \leq 0.05$. This allows us to recover that in the study performed by Rodriguez *et al* [1], where both $A$ and $\alpha$ are small ($A = 0.06$ and $\alpha = 0.02$), the $\mu$ values are hardly affected by the equilibrium term, as noted by the authors themselves. This is also the case in the $CdCr_{1.7}In_{0.3}S_4$ spin glass where $A$ is very small. For the samples where $A$ is not small we have no estimate of its influence on $\mu$, and the only way is to visually appreciate the effect of various ($A$, $\mu$) values on the quality of the scaling. Fortunately, as stated above, $A$ and $\mu$ do not play the same role in the scaling and some additional independent insight on $A$ and $\alpha$ can be derived from a.c. experiments.

---

[20] We assume that the magnetization is the sum of a stationary part and of an aging part. This assumption is made also in reference [1] (to which we compare our results) as well as in many other experimental studies (e.g. reference [2]). A modified scaling form has been proposed recently in a preprint by Kenning, Rodriguez, and Orbach (Cond-mat/0402509): for values of $t/t_w$ as large as 1000, it may be that a third component is involved, which enters into the magnetization in a non-additive way. However, this question is well beyond the scope of our present work, since we focus on the experimental measurement of $\mu$ in the same framework as that used by Rodriguez *et al* [1].

[21] As noted by a referee, the long time behavior of the TRM decays resembles a power law with an exponent which is close to that of the stationary decay. (However, although this seems true for the $CdCr_{1.7}In_{0.3}S_4$ and $Fe_{0.5}Mn_{0.5}TiO_3$ samples, the exponents for the

Au:$Fe_{8\%}$ sample are rather different). The interpretation of this possible connection between short time and long time decay of the TRM remains an open question.

[22] M. Alba, J. Hammann, M. Ocio, Ph Refregier and H. Bouchiat, *J. Appl. Phys.* **61** (8) 3683 (1987)

[23] V. Dupuis, E. Vincent, J-P. Bouchaud, J. Hammann, A. Ito and H. A. Katori, *Phys. Rev. B.* **64** 174204 (2001)

[24] P. E. Jönsson, H. Yoshino and P. Nordblad, *Phys. Rev. Lett.* **89** 097201 (2002)

[25] L. Berthier and J.-P. Bouchaud, *Phys. Rev. B* **66,** 054404 (2002)